\documentclass[%
 amsmath,amssymb,
 aps,
floatfix,
aps,prd,longbibliography,
notitlepage
]{revtex4-1}

\usepackage{graphicx}
\usepackage{dcolumn}
\usepackage{multirow}
\usepackage{natbib}

\newcommand{\BE}{\begin{equation}}
\newcommand{\EE}{\end{equation}}

\newcommand{\BV}{buoyancy }

\begin{document}

\preprint{APS/123-QED}

\title{Observations of the stratorotational instability in rotating concentric cylinders}

\author{Ruy Ibanez}
 \affiliation{Center for Nonlinear Dynamics and Physics Department, University of Texas at Austin, Austin, TX 78712} 
 \email{swinney@chaos.utexas.edu}

\author{Harry L. Swinney}
\affiliation{Center for Nonlinear Dynamics and Physics Department, University of Texas at Austin, Austin, TX 78712} 
 \email{swinney@chaos.utexas.edu}
 
\author{Bruce Rodenborn}%
\email{bruce.rodenborn@centre.edu}
\affiliation{Physics Program, Centre College, Danville, KY 40422}
\date{\today}

\begin{abstract}
We study the stability of density stratified flow  between co-rotating vertical cylinders with rotation rates $\Omega_o < \Omega_i$ and radius ratio $r_i/r_o=0.877$, where subscripts $o$ and $i$  refer to the outer and  inner cylinders. Just as in stellar and planetary accretion disks, the flow has rotation, anticyclonic shear, and a stabilizing density gradient parallel to the rotation axis.   The primary instability of the laminar state leads not to axisymmetric Taylor vortex flow but to the non-axisymmetric {\it stratorotational instability} (SRI), so named by Shalybkov and R\"udiger (2005).  The present work extends the range of Reynolds numbers and buoyancy frequencies ($N=\sqrt{(-g/\rho)(\partial \rho/\partial z)}$) examined in the previous experiments by Boubnov and Hopfinger (1997) and Le Bars and Le Gal (2007). Our observations reveal that the axial wavelength of the SRI instability increases nearly linearly with Froude number,  $Fr= \Omega_i/N$.  For small outer cylinder Reynolds number, the SRI occurs for inner inner Reynolds number larger than for the axisymmetric Taylor vortex flow (i.e., the SRI is more stable). For somewhat larger outer Reynolds numbers the SRI occurs for  smaller inner Reynolds numbers than Taylor vortex flow and even below the Rayleigh stability line for an inviscid fluid. Shalybkov and R\"udiger (2005) proposed that the laminar state of a stably stratified rotating shear flow should be stable for $\Omega_o/ \Omega_i > r_i/r_o$, but we find that this stability criterion is violated for $N$ sufficiently large; however, the destabilizing effect of the density stratification diminishes as the Reynolds number increases. At large Reynolds number the primary instability leads not to the SRI but to a previously unreported nonperiodic state that mixes the fluid. 

\end{abstract}

\maketitle


\section{Introduction}

Stellar and planetary accretion disks that have Keplerian or quasi-Keplerian velocity profiles $V_\phi(r)\propto r^{-1/2}$ are a canonical example of a rotating  flow with anticyclonic shear. Physical and experimental models of accretion disks focus on understanding the mechanisms by which angular momentum is transported radially outward to allow collapse of the central object and planetary bodies. These mechanisms broadly include magneto-hydrodynamic and purely hydrodynamic processes \cite{balbus_hawley_1998}, but studies of concentric rotating cylinder systems have led to a debate whether purely hydrodynamic processes make significant contributions to the angular momentum transport \cite{burin_et_al_2010, paoletti_et_al_2012}. However, many accretion disks include cold regions where the effects of magnetic fields are minimal and the hydrodynamic contribution to the angular momentum transport via turbulent diffusion may be important \cite{shakura_sunyaev_1973, marcus_et_al_2015}.  Understanding the hydrodynamic stability of rotating shear flows is critical to answering this question, and an axial density gradient affects this stability.  
Molemaker et al. \cite{molemaker_et_al_2001} showed that the introduction of weak stratification in a rotating shear flow leads to the growth of a linear, purely hydrodynamic instability. Dubrulle et al. \cite{dubrulle_et_al_2005a} extended this work into the astrophysical context as another possible mechanism for angular momentum transport in accretion disks.

We examine instability of flow between vertical concentric co-rotating cylinders with a stabilizing axial density gradient. 
The parameters defining the problem and their values in our experiment are given in Table I.  

\begin{table}[ht]
\caption{Experimental Parameters}
\begin{tabular}{|c|c|c|}
\hline
\multicolumn{3}{|c|}{\multirow{2}{*}{Concentric Rotating Cylinder System}}\\
\multicolumn{3}{|c|}{ }\\
\hline
inner radius& outer radius& height\\
$r_i=4.218$ cm&$r_i=4.811$ cm & $h=25.6$ cm \\
\hline
radius ratio & gap width & aspect ratio \\
$\eta=r_{i}/ r_{o}=0.877$ &  $d=r_o-r_i=0.593$ cm &$\Gamma=h/d=43.4 $\\
\hline
 inner rotation rate& outer rotation rate & kinematic viscosity\\
$\,\Omega_i$& $\Omega_o$ &$\nu$ \\
\hline
\multicolumn{3}{|c|}{\multirow{2}{*}{Control}}\\
\multicolumn{3}{|c|}{ }\\
\hline
$\,$inner Reynolds number$\,$ &$\,$ outer Reynolds number $\,$&$\,$ bouyancy frequency$\,$\\
$Re_i=(\Omega_i r_i)\,d/\nu$&$Re_o=(\Omega_o r_o)\,d/\nu$&$N=\sqrt{-{g\over \rho} {d\rho\over dz}}$\\
\hline
rotation ratio&Froude number&$q$ parameter \\
 $\mu=\Omega_o/\Omega_i$&$Fr =\Omega_i/N$ &${\Omega_o/\Omega_i}=\eta^q$\\
\hline
\end{tabular}
\end{table}

Our work extends  the Reynolds number range of the laboratory experiments of Boubnov and Hopfinger \cite{boubnov_1997} and Le Bars and Le Gal \cite{lebars_legal_2007} and extends the range of density gradients to about twice that studied in previous experiments \cite{boubnov_1997, lebars_legal_2007}.

This paper is organized as follows: Section \ref{sec:background} describes previous experimental and theoretical work, and Section \ref{sec:methods} describes our experiment and data analysis. Section \ref{sec:results} presents our results for the onset and characterization of the SRI instability and presents observations of a new flow state that mixes the background stratification.  Section \ref{sec:discussion} discusses the results and considers the relevance of our experiments as a model of an accretion disk.

 \section{Background}\label{sec:background}
 
Experimental evidence for hydrodynamic angular momentum transport in rotating shear flows comes primarily from studies of turbulent flow of a fluid of uniform density contained between rotating cylinders. High Reynolds number ($Re\sim 10^6$) Taylor-Couette experiments have found angular momentum transport at rates necessary to allow gravitational collapse within a disk if extrapolated to astrophysical scales \cite{paoletti_et_al_2012}. However, finite size effects from the top and bottom lids on the cylinders confound the results, making these experiments  a poor model for accretion disks \cite{burin_et_al_2010, avila_2012, nordsiek_et_al_2015, edulund_ji_2015}.

Dubrulle et al. \cite{dubrulle_et_al_2005a} emphasize that rotation and shear are not the only important hydrodynamic properties of accretion disks. Protostellar accretion disks are gravitationally bound, and  photo-evaporation by the protostar heats the outer layers of a disk, causing axial variations in density \cite{balbus_hawley_1998}. The effects of stratification in accretion disks are often neglected in theory and in experiments under the assumption that stratification stabilizes the fluid.  This assumption was  supported by early stratified Taylor-Couette experiments \cite{withjack_chen_1974, Boubnov_1995qp, Caton_2000hh}, which found that a stable axial density stratification stabilizes the flow with respect to centrifugal instabilities. However, a 1997 study by Boubnov and Hopfinger \cite{boubnov_1997} found that for co-rotating cylinders  a gravitationally stable density variation can be destabilizing. 

An analysis by Molemaker et al. \cite{molemaker_et_al_2001} shows that when both cylinders are rotating,  stratified Taylor-Couette flow destabilizes through the growth of a linear instability. Dubrulle et al. \cite{dubrulle_et_al_2005a} used both linear stability analysis and a WKB method to show that the instability can be present in accretion disks and therefore may contribute to angular momentum transport in these disks. Shalybkov and R\"udiger \cite{shalybkov_rudiger_2005} followed with a theoretical study that named this instability the {\it stratorotational instability} (SRI). 

For a fluid of uniform density the well-known Rayleigh criterion for stability of inviscid rotating shear flows is \cite{lord_rayleigh_1880}
\BE
{d(\Omega r^2)^2\over dr} >0,
\EE
where $\Omega(r)$ is the radially varying rotation rate of the shear flow and $r$ is the radial coordinate.  For the concentric cylinder system the Rayleigh criterion becomes $\mu > \eta^2$, where $\mu$ is the ratio of the cylinder rotation rates, $\Omega_o/\Omega_i$, and $\eta$ is the ratio of the cylinder radii, $r_i/r_o$ \cite{chandresekhar_1961}. Shalybkov and R\"udiger \cite{shalybkov_rudiger_2005} proposed an analogous criterion for stability in a stratified rotating shear flow,
\BE \label{eq:stratstab}
{d(\Omega r)^2\over dr}>0,
\EE
which can be written more compactly as  $\mu > \eta$. 

The first detailed experiment on the SRI in stratified Taylor-Couette flow was by Le Bars and Le Gal \cite{lebars_legal_2007}, who confirmed the $\mu>\eta$ stability criterion over the range of Reynolds numbers  they explored (cf. Table I). The experiments of Le Bars and Le Gal and the simulations of Shalybkov and R\"udiger \cite{shalybkov_rudiger_2005} examined the inner Reynolds number range $0<Re_i< 1200$. The measurements in Le Bars and Le Gal were for the Froude number $Fr=0.5$. Dubrulle et al. \cite{dubrulle_et_al_2005a} estimate that $Fr\sim 3$  for an accretion disk;  they found that  large $N$ stabilizes flow beyond the stability onset of Taylor vortex flow (TVF), while for small $N$ the flow becomes unstable below the onset of  the TVF instability \cite{dubrulle_et_al_2005a}.  

This paper expands the range of density gradients examined in experiments by a factor of two. We also provide a detailed study of the SRI at Reynolds numbers $Re_i>10,000$, for which Boubnov and Hopfinger \cite{boubnov_1997} provided onset data but did not describe the flow states.

\begin{figure}[ht]
  \centering
    \includegraphics[width=\columnwidth]{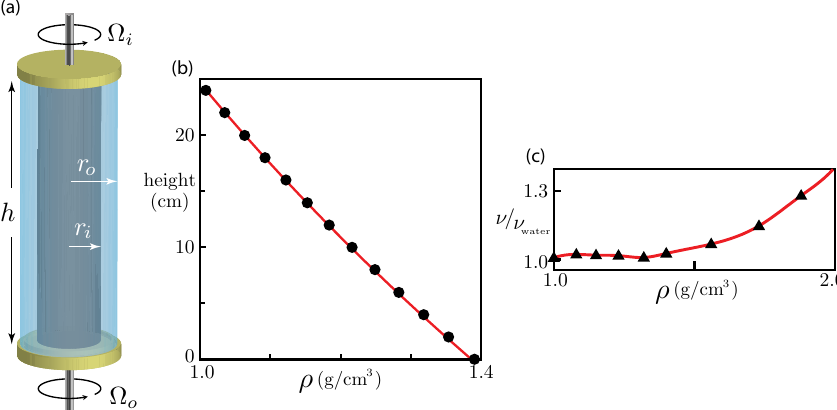}
    \caption{(a) A schematic diagram of the experimental system. (b) Density as a function of height for $N$ = $3.14$ s$^{-1}$. The black circles are measurements and the solid red curve is a fit of the data to an exponential. (c) Viscosity of an aqueous solution of sodium polytungstate salt, as a function of density. Black triangles are the manufacturer's data for the kinematic viscosity $\nu$ at 25 $^\circ$C \cite{geoliquids} normalized by the viscosity of water $\nu_{\rm water}$.  The solid red curve is a cubic spline fit to the manufacturer's data.}\label{fig:schematic}
\end{figure}

\section{Methods}\label{sec:methods}

The parameters for our Taylor-Couette system (cf. Fig. \ref{fig:schematic}(a)) are given in Table I. The cylinders rotate independently, driven by computer-controlled stepper motors.


\subsection{Stratification}
The data presented here are for three fixed values of $N$: $1.57$, $3.14$, and $4.71$ s$^{-1}$. Previous experiments achieved a density gradient using  NaCl solutions, which have a maximum density of 1200 kg/m$^3$ (hence a maximum density range of 20\%), while the present experiments use sodium polytungstate salt (Na$_6$H$_2$W$_{12}$O$_4$) solutions, which have a maximum density greater than 3000 kg/m$^3$ \cite{geoliquids}, making it possible to vary the density of the salt solutions by a factor of three.

Large $N$ is not accessible for large aspect ratios ($h>>d$) when using NaCl solutions, which have a maximum density of 1200 kg/m$^3$ (hence a maximum density range of 20\%).  Our experiments use sodium polytungstate salt (Na$_6$H$_2$W$_{12}$O$_4$) solutions, which have a maximum density greater than 3000 kg/m$^3$ \cite{geoliquids}; thus the density of the salt solutions can vary by a factor of three.  We used a maximum density variation of a factor of two because at higher densities the viscosity increases rapidly with density.

We use two computer-controlled syringe pumps to create a stable vertical stratification in the gap. One syringe contains fresh water and the other contains an aqueous solution of sodium polytungstate along with a small amount ($<1\%$) of Kalliroscope flakes for visualization. A Matlab algorithm calculates the pump rates necessary to create an exponentially varying density $\rho(z)$ (cf. Fig. \ref{fig:schematic}(b)).  We use an exponential density gradient to keep the \BV frequency, $N$, constant. 

We determine the density stratification  by removing approximately 1 ml of fluid at height increments of 2 cm using a syringe with a long needle mounted to a translation stage driven by a stepper motor. The extracted fluid is analyzed with an Anton-Paar density meter, which is accurate to $\pm 1$ kg/m$^{3}$.  The fluid sample is then re-injected into the system to minimize the effects of the measurement. 

The viscosity of sodium polytungstate aqueous solutions depends on the salt concentration, and therefore the Reynolds number varies as a function of height. We obtain the viscosity as a function of height using a cubic spline fit  to the viscosity values given by the manufacturer \cite{geoliquids} (cf. Fig. \ref{fig:schematic}(c)). We compute $N$ using the average kinematic viscosity value in the middle third of the system, which is the region that we analyze. Over this region the viscosity varies at most $\pm2.2$\% for the largest stratification we use, and the variation with height is less than 1\% for the other stratifications. 

The \BV frequency is calculated by fitting an exponential curve to the density measurements. An example density profile is given in Fig. \ref{fig:schematic}(b). The density is accurately exponential except near the ends of the annulus, where the deviations from an exponential can become pronounced due to Ekman cells. However, these deviations  do not significantly affect the stratification in the central region that we analyze.

\subsection{Determining Onset of Instability}

After preparing the background stratification, we determine the instability boundary by accelerating the cylinders to solid body rotation ($\mu = 1$) and then decreasing $Re_o$  while $Re_i$ is held fixed. The instability onset is determined to within 1\% of $Re_o$ by stepping back and forth through the instability boundary in a binary search. Measurements of the axial wavelength of the stratorotational state as a function of Froude number, $Fr\equiv \Omega_i/N$, are made by varying $\Omega_i$ for each of the three values of $N$. 
\begin{figure}[ht]
  \centering
    \includegraphics[width=.65\columnwidth]{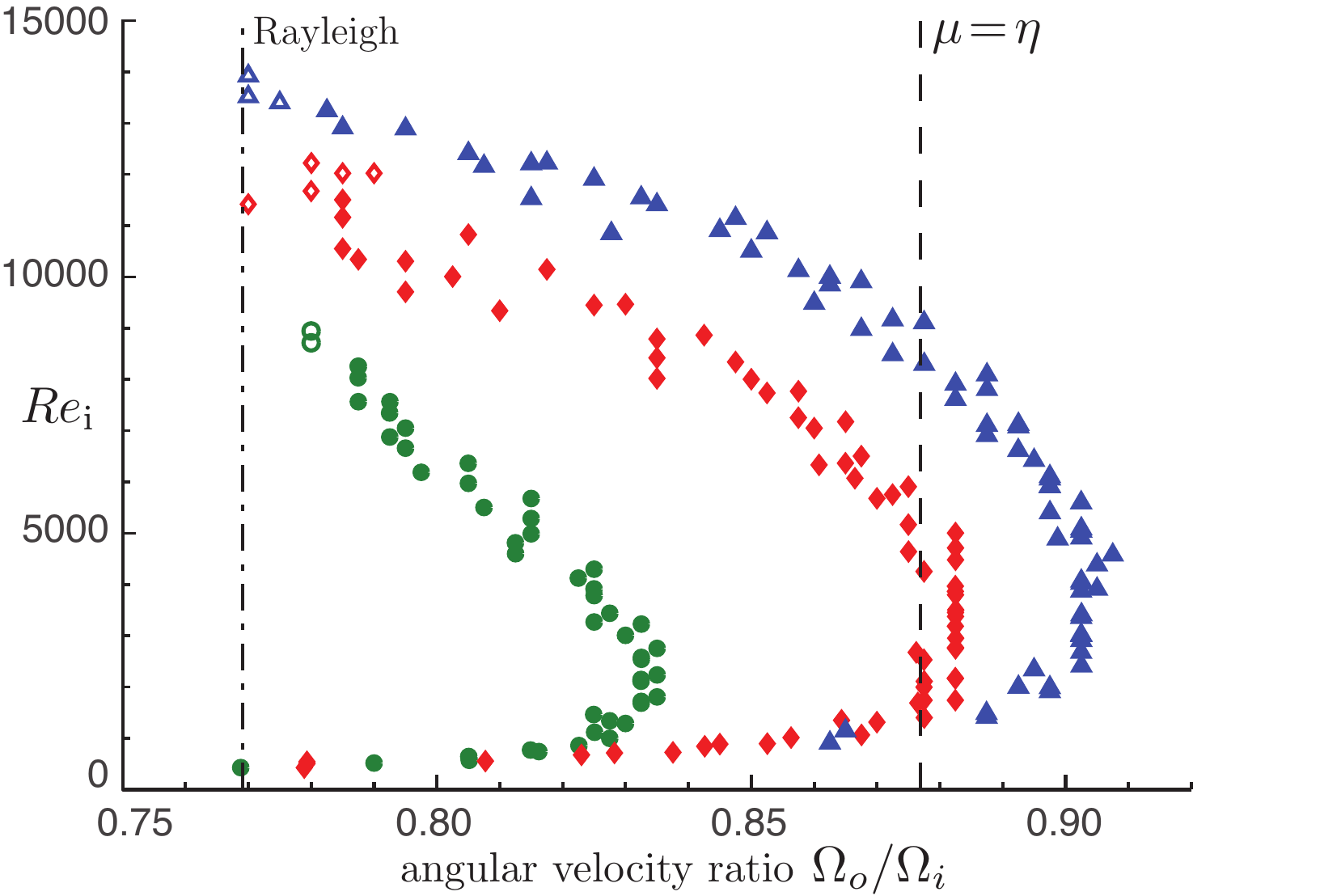} 
\caption{ Observations of the SRI onset for increasing anticyclonic shear as $\mu=\Omega_o/\Omega_i$ was decreased from solid body rotation  ($\mu=1$) and $Re_i$ was held fixed (green symbols, $N=1.57$ s$^{-1}$; red, $N=3.14$ s$^{-1}$; blue, $N=4.71$ s$^{-1}$).  Rayleigh's stability criterion for inviscid unstratified flow predicts stability to the right of the vertical dot-dash $\mu=\eta^2$ line \cite{lord_rayleigh_1880}, while for viscous stratified flow Shalybkov and R\"udiger \cite{shalybkov_rudiger_2005} predict stability to the right of the dashed $\mu=\eta$ line.}
\label{fig:onset}
\end{figure}

\begin{figure}[ht]
    \includegraphics[width=\columnwidth]{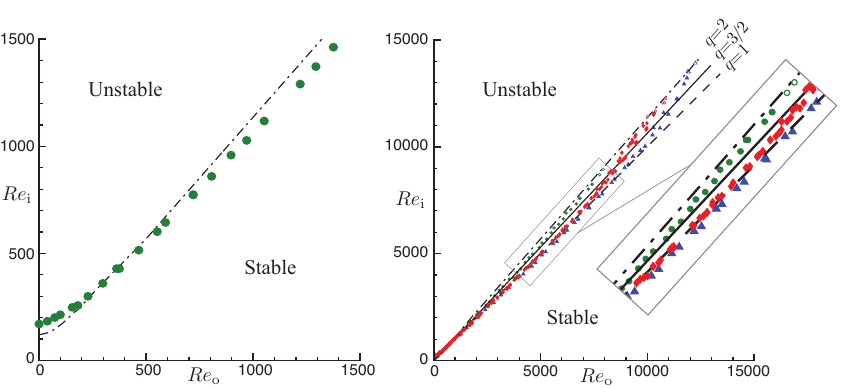}
    \caption{(a) These data for low Reynolds numbers $Re_i$ and $Re_o$ show that a vertical density gradient stabilizes the laminar flow compared with the curve for the unstratified Taylor vortex flow instability (dot-dash curve), but for $Re_i$ and $Re_o$ above about 350 a density gradient destabilizes the laminar flow before the onset of Taylor vortex flow.  
(b) Data for $Re_i$ and $Re_o$ an order of magnitude larger than in (a) show that the laminar flow is destabilized more for a larger density gradient, but the effect of stratification decreases at higher Reynolds numbers (green symbols, $N=1.57$ s$^{-1}$; red, $N=3.14$ s$^{-1}$;  blue, $N=4.71$ s$^{-1}$). The open symbols correspond to a nonperiodic state that mixes the fluid. The Taylor stability line \cite{taylor_1923} is the dot-dashed curve (labeled $q=2$ using the notation of Balbus and Hawley \cite{balbus_hawley_1998}), which is indistinguishable from the Rayleigh line at high Reynolds numbers.  The Keplerian velocity profile is the  solid line ($q=3/2$), and the dashed line ($q=1$) is the Shalybkov and R\"udiger prediction for instability onset. The inset enlarges the region where the data cross the $\mu=\eta$ line.  The perpendicular distance in Reynolds number between the $q=1$ and $q=2$ lines is at most 600.}
\label{fig:rei_reo}
\end{figure}

As noted by Le Bars and Le Gal \cite{lebars_legal_2007}, the SRI does not mix the fluid even if the stratorotational state is observed for very long times. We conducted extensive measurements of the onset of the SRI for periods even longer than one day and found that the density stratification did not change significantly, except at the ends, which showed mixing due to Ekman circulation. The Ekman cells were typically 3 cm or less high, but at the largest $Re_i$, 14,000, the Ekman cells were 7 cm high.

\subsection{Spatial and Temporal Frequencies}
We analyze the spatiotemporal characteristics of the pattern by capturing images using an Edgertronic high speed monochrome camera and/or a Nikon D200 digital still camera. We analyze the time series intensity data for a narrow vertical strip in the central region of the digital image. The power spectral density is found using the fast Fourier transform function in Matlab  (cf. Fig. \ref{fig:spectrum}). 

We determine the axial wavelength by measuring the distance in pixels between successive vortex pairs in a still image (cf. Fig. \ref{fig:lambda}).    The length in pixels is converted into centimeters by imaging a ruler near the region being analyzed. 

\section{ Results }\label{sec:results}
\subsection{Comparison with Previous Work}
		
Shalybkov and R\"udiger \cite{shalybkov_rudiger_2005}  predicted that the laminar flow should be stable if $\mu>\eta$, and Le Bars and Le Gal's observations were consistent with this prediction for the \BV frequencies  and Reynolds numbers ($Re_i \leq 1200$) in their experiments. Our measurements agree with those of Le Bars and Le Gal for the the range of \BV frequencies and Reynolds numbers that they examined. However, our measurements with $Re_i$ ranging up to 14,000 reveal that for $N>3.1$ s$^{-1}$ the $\mu>\eta$ theoretical stability criterion of Shalybkov and R\"udiger does not hold, as Fig. \ref{fig:onset} illustrates.  This observation of the failure of the stability criterion is in accord with later numerical simulations by Shalybkov and R\"udiger \cite{Rudiger:2009zf}.

\subsection{Onset of the SRI }
The SRI onset data in Fig. \ref{fig:onset} are replotted in Fig. \ref{fig:rei_reo} for comparison with the classic TVF instability of an unstratified fluid. The nonaxisymmetric SRI and axisymmetric TVF instability onset curves fall close to one another, even for large $N$.  At small $Re_o$ a density gradient stabilizes the stratified fluid compared to the TVF, but for $Re_o > 350$  the SRI precedes the TVF instability (Fig. \ref{fig:rei_reo}).

The destabilizing effects of stratification increase with Reynolds number until $Re_o\gtrsim$ 7500, at which point the onset of instability approaches the Taylor line (Fig. \ref{fig:rei_reo}(b) inset). The effects of stratification are diminished by global rotation and are small at high Reynolds number. The SRI occurs both above and below the line corresponding to a Keplerian velocity profile ($q=3/2$), and the SRI also occurs both above and below the $\mu=\eta$ ($q=1$) line, where we have used the $q$ parameter of Balbus and Hawley\cite{balbus_hawley_1998}. Thus, the onset of the SRI does not depend on a particular $q$-value, but is a function of both stratification and Reynolds number.

\begin{figure}[ht]
  \centering
    \includegraphics[width=\columnwidth]{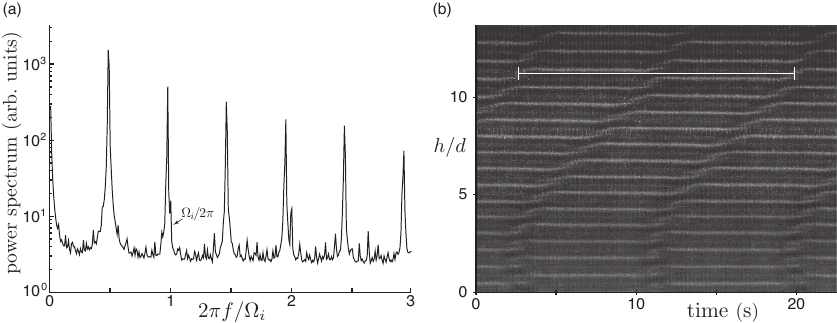}
    \caption{(a) This power spectrum of a movie pixel intensity time series shows that the SRI state is simply periodic; the six harmonics yield a fundamental frequency $2\pi f/\Omega_i=0.4878$, corresponding to a period of 17.23 s.  The movie was taken with $Re_i$ about 1\% above onset of the SRI for
    $N=1.57$ s$^{-1}$,  $\Omega_i/2\pi=0.119$ Hz, and $\Omega_o = 0$ ($Re_i=169$, the lowest data point in Fig. \ref{fig:rei_reo}(a)). (b) An image sequence from a movie of the flow state in panel (a). Strips of 16 pixels from the same region in each frame are stitched together to show the propagation of the oscillatory disturbance. The white bar shows the period corresponding to the fundamental frequency in panel (a). Note that the pattern repeats after two waves pass a given point.  }
\label{fig:spectrum}
\end{figure}	
\subsection{Frequency of SRI}\label{sec:TempFreq}

We determine the frequency of the SRI from a spectral analysis of time series of movie pixel intensities, as illustrated by the power spectral density graph in Fig. \ref{fig:spectrum}(a).  The SRI frequency is nearly equal to the average rotation rate of the inner and outer cylinders, as previously observed by Le Bars and Le Gal \cite{lebars_legal_2007}. For example, in  Fig. \ref{fig:spectrum}(a) the second harmonic of the SRI frequency is only 2.4\% smaller than $\Omega_i /2\pi$, which is twice the average frequency of the cylinders for these data that were obtained with $\Omega_o = 0$.  

\subsection{Axial Wavelength}
\begin{figure} [ht]
\begin{flushleft}
\includegraphics[width=\columnwidth]{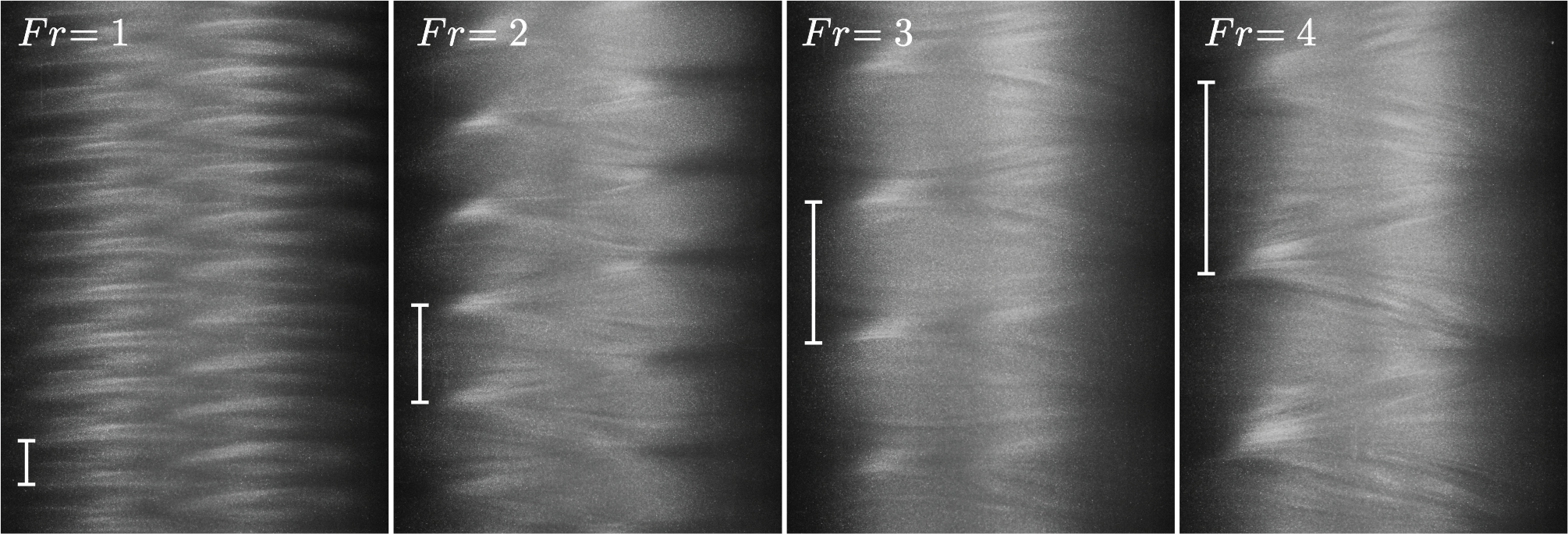}
\end{flushleft}
\caption{Images of the SRI pattern for four values of Froude number $Fr=\Omega_i/N$ with $N=1.57$ s$^{-1}$. The  wavelengths determined from the images are plotted in Fig. \ref{fig:lambda}. The length of the smallest scale bar is $\lambda=1.81\,d$, which is 1.07 cm.}
\label{fig:lambdaims}
\end{figure}

Figure \ref{fig:lambdaims} illustrates that the SRI axial wavelength $\lambda$ increases linearly with $Fr$, as predicted by the inviscid analysis of Molemaker et al. \cite{molemaker_et_al_2001},  who found 
\begin{equation}
\lambda/d=(2\pi Ro\sqrt{Ro+1})\,Fr,
\label{eq:lambda_predicted}
\end{equation}
where the Rossby number, $Ro$, defined by their Eq. (4), is approximately 0.4 in our work. The prediction of Molemaker et al. \cite{molemaker_et_al_2001} for the slope is compared with our measurements in the caption of Fig. \ref{fig:lambdaims}. Beyond the onset of SRI, we find little change in the wavelength.

\begin{figure}[ht]
  \centering
    \includegraphics[width=.65\columnwidth]{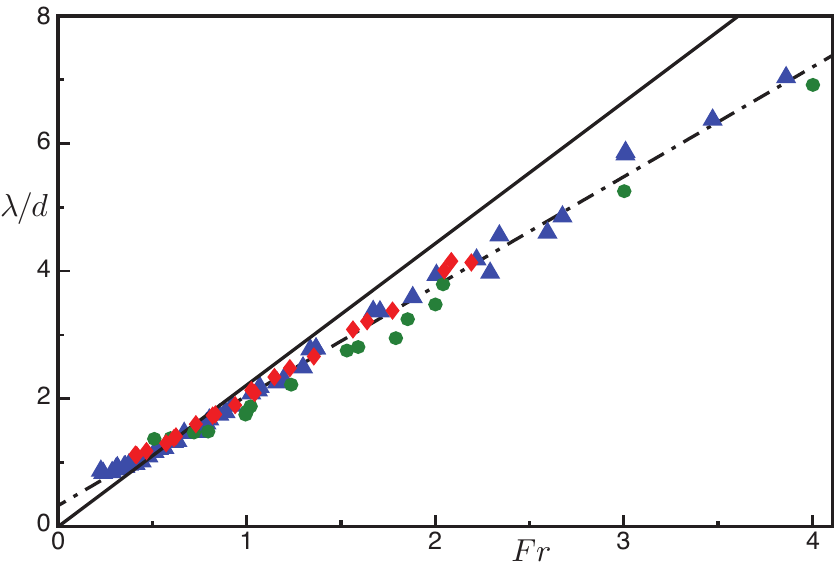}
    \caption{Axial wavelength $\lambda$ at the onset of the SRI, normalized by the gap width $d$, as a function of $Fr=\Omega_i/N$.  The symbols are the same as in Figs. 2 and 3 (green symbols $N=1.57$ s$^{-1}$; red, $N=3.14$ s$^{-1}$;  blue, $N=4.71$ s$^{-1}$). 
A linear fit to the data (the black dash-dot line) yields $\text{slope} = 1.72$, which is smaller than the slope value 2.2 predicted by linear inviscid theory (solid line), Eq. \ref{eq:lambda_predicted} \cite{molemaker_et_al_2001}. } 
\label{fig:lambda}
\end{figure}

\subsection{High Reynolds Number}\label{sec:highRe}
	At Reynolds numbers greater than 2000, above the range examined by Le Bars and Le Gal, the instability curves in a plot of $Re_i$ vs $\mu$ turn upward (cf. Fig. \ref{fig:onset}). For $Re_i$ larger than about 5000, the SRI occurs for lower values of $\mu$; thus larger global rotation stabilizes the fluid with respect to SRI (Fig. \ref{fig:onset}). This behavior is not predicted by existent SRI theory and was not reported in the experiments by Boubnov and Hopfinger \cite{boubnov_1997} or in computational studies.		
	
\begin{figure*}[ht]
  \centering
\includegraphics[width=\columnwidth]{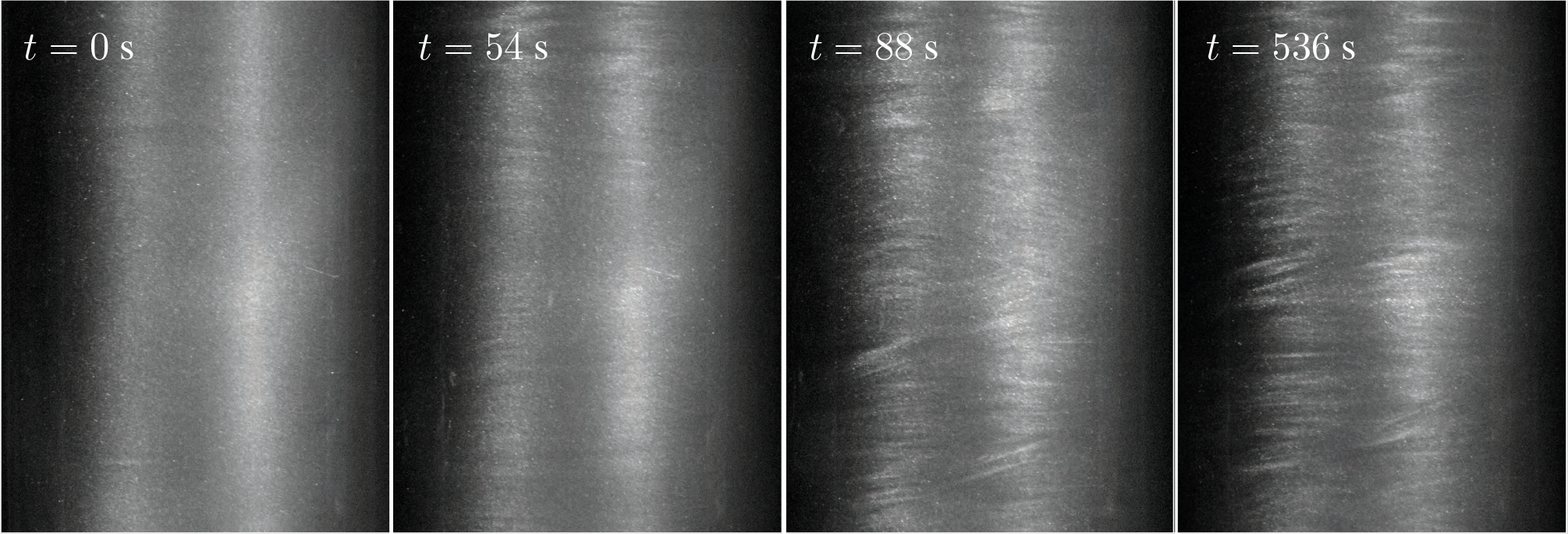}
    \caption{ Images showing the development of a nonperiodic state from laminar flow at $Re_i=9000$ after the ratio of the cylinder rotation rates, $\mu$, was slowly reduced from unity to $\mu = 0.78$  ($N = 1.57$ s$^{-1}$).   The fluid continues to evolve during the nine minutes after the radius ratio was set at $\mu = 0.78$.}
\label{fig:chaos}	
\end{figure*}

\begin{figure}[ht]
  \centering
    \includegraphics[width=.65\columnwidth]{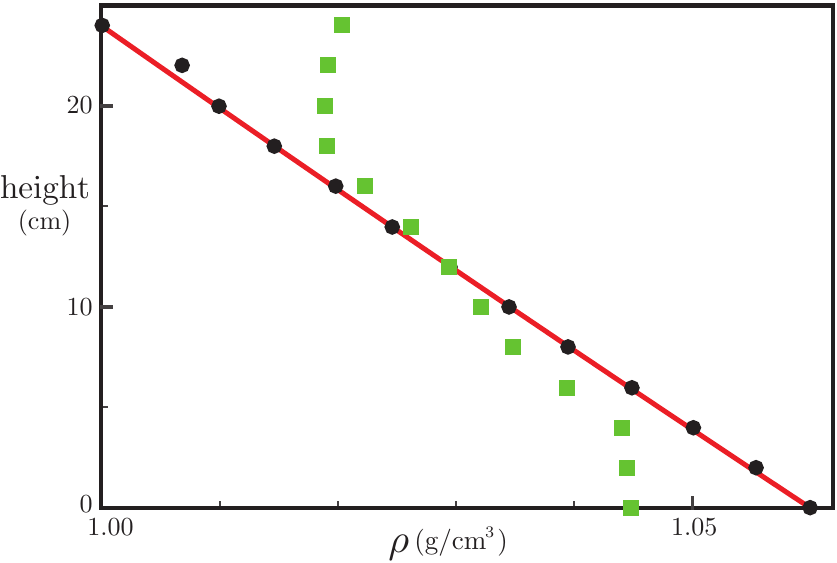}
    \caption{Density measurements (black circles) with an exponential fit (red curve) before the fluid transitions  from the laminar to the nonperiodic state. Green squares show the density 8.9 minutes after the transition to the nonperiodic state (cf. Fig. \ref{fig:chaos}).}
\label{fig:mixing}
\end{figure}

Figures \ref{fig:onset} and \ref{fig:chaos} show that for large $Re_i$, instability of the laminar state leads not to the SRI but to a previously unreported nonperiodic state, which rapidly mixes the fluid. Larger stratification suppresses the onset of this nonperiodic state (Figure \ref{fig:rei_reo}). The characteristic helicoidal modes of SRI are not present in the nonperiodic state, although in addition to the fine scale structure there is large scale structure, as can be seen in  Fig. \ref{fig:chaos}. The mixing in the chaotic state reduces the density stratification, as Fig. \ref{fig:mixing} illustrates.  A spectral analysis of the nonperiodic state reveals the same fundamental frequency and harmonics as the SRI because the large scale structure rotates at the average speed of the two cylinders, but the mixing at small scales increases the background noise level in the power spectra.
 
The nonperiodic state appeared at a fairly well defined value of $\mu$ as it was decreased in small steps ($\Delta \mu\approx 0.005$). It was not possible to determine whether there was hysteresis in the onset of the nonperiodic state because the fluid mixed rapidly once the nonperiodic state appeared. Consequently, we were unable to cross back and forth from the laminar to the nonperiodic state in the same way as  was done in determining the onset of the SRI state.

We examined the criterion of Richard and Zahn \cite{Richard_Zahn_1999}, who define a critical {\it gradient} Reynolds number, $Re^*$, for the onset of subcritical instability in a uniform fluid:
	\begin{equation}
Re^* \equiv{r_m^3\over \nu}{ |\Omega_i-\Omega_o|\over d}\gtrsim 6\times 10^5,
\label{eq:subcrit}
\end{equation}
	where $r_m$ is the mean value of the inner and outer radii. The gradient Reynolds numbers at which we observed the nonperiodic state were $Re^*\geq 6.7\times 10^5$, which satisfies the condition in Eq. (\ref{eq:subcrit}).

\section{ discussion } \label{sec:discussion}

Studies of the stability of a stratified fluid between rotating concentric cylinders have been motivated by the possibility that this system might  provide insight into fluid instability of astrophysical accretion disks. The experiments have used vertical co-rotating cylinders with an axial density gradient. In order to impose an anticyclonic shear as in accretion disks, the outer cylinder is rotated more slowly than the inner cylinder. The experiments have revealed an instability of the laminar flow to the helicoidal ``Stratorotational Instability'' (SRI) \cite{withjack_chen_1974, boubnov_1997, lebars_legal_2007}.  The SRI is an interesting new flow state in the rotating concentric cylinder system, which has been extensively studied for nearly a century following G.I. Taylor's pioneering study \cite{taylor_1923} of instability of   the laminar flow (in a constant-density fluid) to axisymmetric vortices (Taylor vortices).  

The present experiment extends by a factor of two the maximum axial density gradient investigated for the SRI.  At large Reynolds numbers we find  the primary instability is not the SRI but rather a nonperiodic flow that rapidly mixes the fluid (cf. Fig. \ref{fig:onset}). Thus it is unlikely that the SRI is relevant at the far higher Reynolds numbers of protoplanetary disks. However, the effect of axial density stratification on protoplanetary disks may  be important through other instabilities such as the subcritical instability proposed by Marcus et al. \cite{marcus_et_al_2013, marcus_et_al_2015}.

We have examined the onset of the SRI for large Reynolds numbers as well as large density gradients. For the two larger density gradients studied, where $N=3.14$ and 4.71 s$^{-1}$, we have  found a range of $Re_i$ where the instability occurs for $\Omega_o/ \Omega_i > r_i/r_o$ (Fig.~\ref{fig:onset}),  contrary to the prediction of \cite{shalybkov_rudiger_2005} but consistent with later numerical simulations \cite{Rudiger:2009zf}.  

While there is a striking difference between the curves for different $N$ in Fig. \ref{fig:onset}, the dependence of SRI onset on stratification is much less striking in the Taylor-like graph of $Re_i$ vs $Re_o$ (Fig.~\ref{fig:rei_reo}).  The symmetry of the emergent helicoidal SRI differs from axisymmetric Taylor vortex flow (TVF) of an unstratified fluid, yet the SRI and TVF instability curves are surprisingly close. The smallest and largest density gradients differ by a factor of nine, but the corresponding instability curves differ in Reynolds number by at most 9\% (cf. Fig. \ref{fig:rei_reo}). Note also that for small $Re_o$, the onset of the SRI occurs for $Re_i$ larger than for the axisymmetric TVF, whereas for $Re_o\gtrsim 350$ the onset of SRI occurs for  smaller Reynolds numbers than the onset of TVF.  

We find a strong dependence of the axial wavelength of the SRI flow on the Froude number $Fr=\Omega_i/N$, as shown in Fig. \ref{fig:lambda}. The wavelength increases linearly from the smallest measured value (0.88$d$) to the largest ($7d$).  In the latter case only five  wavelengths fit between the Ekman cells at the ends of the annulus; however, given the large density gradient, we presume that end effects do not significantly affect the SRI flow. 

In conclusion, the present study has found that at large $Re_i$ (about 9000 for $N=1.57$ s$^{-1}$ and 14,000 for $N=4.71 s^{-1}$), the instability of the base flow that occurs with increasing anticyclonic shear (decreasing $\Omega_o/\Omega_i$) leads to nonperiodic flow rather than the periodic SRI.  In Taylor-type graphs of $Re_i$ vs $Re_o$, the SRI onset line is close to the Taylor vortex onset line in an unstratified fluid, where the radial dependence of the velocity for the azimuthal flow is close to the Rayleigh profile \cite{lord_rayleigh_1880}, $V_\phi(r)\propto r^{-2}$ ($q=2$, cf. Fig. \ref{fig:rei_reo}), rather than to the approximate Keplerian profile $V_\phi(r)\propto r^{-3/2}$ ($q=3/2$) of protoplanetary disks. 

We thank Hepeng Zhang, Abhay Argarwal, Evander Harris, Shuqi Liu, John Scelzi and Deon Rodenborn for their assistance. This research was supported in part by the Sid W. Richardson Foundation.

\end{document}